\documentclass{article}
\usepackage[a4paper=true,hypertex]{hyperref}
\usepackage{epsfig,amssymb,amsmath}
\author{ Diego Harari, Silvia Mollerach, Esteban Roulet\\
{\it CONICET, Centro At\'omico Bariloche,}\\
{\it Av. Bustillo 9500 (8400) Argentina.}}

\title{Anisotropies of ultra-high energy cosmic rays \\
diffusing from extragalactic sources }

\begin{document}

\maketitle

\begin{abstract}
 We obtain the dipolar anisotropies in the arrival directions of ultra-high energy cosmic rays diffusing from nearby extragalactic sources. We discuss both the energy regime of spatial diffusion and the quasi-rectilinear one leading to just angular diffusion at higher energies. We obtain analytic results for the anisotropies from a single source  which are validated using two different numerical simulations. 
For a scenario with a few sources  in the local supercluster (with the closest source at a typical distance of few to tens of Mpc), we discuss the possible transition between the case in which the anisotropies are dominated by a few sources at energies below  few EeV towards the regime in which  many sources contribute at higher energies. The effect of a non-isotropic source distribution  is also discussed, showing that it can significantly affect the observed dipole.

\end{abstract}

\section{Introduction}

The actual sources of the cosmic rays (CRs) are still unknown, although it is believed that the vast majority of those observed with energies $E<0.1$~EeV (where ${\rm EeV}\equiv 10^{18}$~eV) are of galactic origin, probably accelerated in supernova explosions, while those with energies above few EeV are most likely of extra-galactic origin, probably accelerated in active galaxies or gamma ray bursts. The precise energy at which the transition between galactic and extra-galactic CRs takes place is under debate, with different proposals locating it near the second-knee (a steepening of the spectrum at $\sim 0.1$~EeV) or near the ankle (a flattening of the spectrum observed at $\sim 4$~EeV). 

The main observables available to infer the CR properties are the energy spectrum, the composition indicators, and the anisotropies in the arrival directions measured at different angular scales.  In particular, the changes in the slope of the spectrum just mentioned  could for instance be indicating a change in the propagation properties (such as an enhanced escape from the Galaxy starting at the second knee) or a change in the source population (such as the dominance of the extra-galactic component above the ankle), and some combination of these  can also take place at intermediate energies. In addition,  other effects such as the impact of pair production energy losses on extragalactic protons interacting with the CMB or diffusion effects from nearby sources  can also contribute in shaping the spectrum at EeV energies.

Regarding the measured composition, the Kascade-Grande surface detector experiment \cite{kg13} suggests a predominantly heavy composition just below 0.1~EeV, as would be expected for a galactic component in scenarios in which the acceleration or the confinement of  CRs in the Galaxy depend on the rigidity of the particles, so that galactic protons start to be suppressed at the observed steepening at the knee (for $E\sim 5\times 10^{15}$~eV), while Fe nuclei get suppressed only at an energy 26 times larger, which is close to the location of the second-knee. A puzzling fact is that the  maxima in the air shower development as well as the shower maxima fluctuations,  determined using the fluorescence technique \cite{augerxmax,hiresxmax}, suggest that at EeV energies  the composition is predominantly light. 
However, a proton component at EeV energies can hardly be of galactic origin, since besides the difficulty of devising appropriate galactic sources that could reach such high rigidities, the galactic protons would be expected to give rise to strong anisotropies towards the direction of the galactic plane, contributing to both a dipolar and a quadrupolar component in large scale anisotropy searches, which are not observed \cite{augerls}.
Actually, the large scale anisotropies in right ascension reported by Auger are below $\sim 2$\% at EeV energies, and show a marginally significant indication of a transition from a direction near $RA\simeq 270^ \circ$ below 1~EeV, which is consistent with the direction of the galactic center, towards directions near $RA\simeq 100^ \circ$ above the ankle energy, with an amplitude increasing to the several percent level near 10~EeV.

In this paper we want to consider in detail the predictions for the anisotropies produced by a nearby extragalactic source in the case in which CRs diffuse due to the presence of sizeable turbulent magnetic fields in the  Local Supercluster, so that spatial diffusion is relevant, having in mind that this may significantly contribute to the CR anisotropies at EeV energies.
 Some studies of this kind were performed in the past \cite{anidif}, and here we improve on the treatment of the diffusion effects, adopting a more accurate energy dependence for the diffusion coefficients and we obtain a detailed matching between the diffusive and quasi-rectilinear regimes, finding analytic fits to the results and validating them with different numerical simulations of CR trajectories in turbulent fields. We also discuss the possible transition between a few sources contributing at low energies to a regime were many sources contribute at higher energies, considering also the effects of their possibly non-isotropic spatial distribution. As by-products of this work we provide in Section~3 detailed fits to the diffusion coefficients as a function of the energy for Kolmogorov and Kraichnan turbulence, covering the transition between the resonant and non-resonant regimes. We also provide in Appendix~II simple fits to the proton attenuation lengths which allow for an analytic treatment of the effects of proton energy losses.

\section{Turbulent magnetic fields and diffusive propagation}

Turbulent magnetic fields may be produced in the Universe from the evolution of primordial seeds affected by the process of structure formation. This typically leads to magnetic fields with strength correlated with the matter density ($B\propto \rho^{2/3}$ due to the flux conservation during the collapse), and hence being enhanced in dense regions such as superclusters while suppressed in the voids. Extragalactic magnetic fields may also result from galactic outflows, in which galactic B fields are transported by winds into the intra-cluster medium. Although magnetic fields with $\mu$G strengths have been measured in cluster cores, at supercluster scales they are expected to be smaller, and  values from nG up to $\sim 100$~nG have been considered, adopting for their coherence length typical values of order 0.1--1~Mpc \cite{bsims}. 

We will start by considering individual CR sources in the local supercluster, at  distances $r_s<100$~Mpc, assuming for simplicity that  a uniform isotropic turbulent magnetic field is present within the diffusion region. The field will be characterized by a root mean square strength $B=\sqrt{\langle B^2(x)\rangle}$. An important property is the distribution of the magnetic energy density $w$ on different length scales, usually described adopting a power law in Fourier space $w(k)\propto k^{-m}$. For instance, a Kolmogorov spectrum of turbulence corresponds to $m=5/3$, and is obtained when the initial energy is injected at the maximum scale $L_{max}$  and is then transferred to lower scales by wave interactions until it dissipates at the scale $L_{min}$. Alternatively, the Kraichnan spectrum with $m=3/2$ is expected to result in the case of MHD waves. 

Several  magnetic field properties are discussed in the Appendix I. In particular the coherence length of the field $l_c$, defined as in \cite{ac99,ha02}, satisfies
\begin{equation}
l_c=\frac{L_{max}}{2}\frac{m-1}{m}\frac{1-(L_{min}/L_{max})^m}{1-(L_{min}/L_{max})^{m-1}}
\label{lc}
\end{equation}
where $L_{min}$ and $L_{max}$ are the minimum and maximum scales of the turbulence spectrum. Note that in the case in which $L_{min}\ll L_{max}$, as we will consider in the following, one has $l_c\simeq L_{max}/5$ for Kolmogorov, while $l_c\simeq L_{max}/6$ for Kraichnan turbulence.
An effective Larmor radius can be introduced as
\begin{equation}
r_L=\frac{E}{ZeB}\simeq 1.1 \frac{E/{\rm EeV}}{Z\,B/{\rm nG}}\,{\rm Mpc},
\end{equation}
with $Ze$ the particle charge.
A crucial quantity to characterize the particle diffusion is the critical energy $E_c$, defined such that $r_L(E_c)=l_c$ and hence given by

\begin{equation}
E_c=Z e B l_c \simeq 0.9 Z\,\frac{B}{\rm nG}\frac{l_c}{\rm Mpc}\,{\rm EeV}.
\end{equation}
This critical energy separates the regimes of resonant diffusion at low energies, in which particles have large deflections induced by their interactions with the $B$ field modes with scales comparable to the Larmor radius, and the non-resonant regime at high energies in which the deflections after traversing a distance $l_c$ are small, typically of order $\delta\simeq l_c/r_L$.

For isotropic diffusion the particle flux is given by ${\vec J}=-D{\vec \nabla} n$, with $n$ the particle density and $D$ the diffusion coefficient, and hence the average distance from the sources $r(t)$ after CRs travel for a time $t$  satisfies $\langle  r(t)^2\rangle = 6Dt$.  
The diffusion length $l_D\equiv 3D/c$ characterizes the distance after which the deflection of the particles is $\simeq 1$~rad. In particular for $E\ll E_c$ one has that $l_D\simeq a_L l_c(E/E_c)^\alpha$, with $\alpha\equiv 2-m$, while for $E\gg E_c$ one has $l_D\simeq a_H l_c(E/E_c)^2$, since in the latter regime one needs to traverse $N\simeq l_D/l_c$ coherent domains to have a total deflection $\delta\simeq 1$~rad, where $\delta\simeq \sqrt{N}(l_c/r_L)$ results from the random angular diffusion of the CR trajectory. The low and high energy coefficients $a_{L,H}$ are of order unity and will be discussed in Section~4, together with a detailed fit to the diffusion length in the transition region at intermediate energies. As long as the source distance is much larger than $l_D$, spatial diffusion of the CR particles will take place. At sufficiently large energies $l_D$ will become larger than $r_s$ and one will enter the quasi-rectilinear regime in which the total deflection of the particles arriving from the source is less than 1~rad, and hence only some angular diffusion will occur but not the spatial diffusion.  The onset of the quasi-rectilinear regime corresponds to the condition $r_s\simeq l_D$, and hence this happens for $E>E_{rect}\equiv E_c\sqrt{r_s/l_c}$ (where we assumed that $E_{rect}>E_c$ so that $D\propto E^2$, which is indeed the case if $r_s\gg l_c$). 

The extragalactic CR diffusion can modify the spectrum of the particles reaching the Earth, specially at low  energies ($E/Z<{\rm EeV}$) due to a magnetic horizon effect \cite{le04,be06,gl07,mo13}, and can also be crucial in the determination of the anisotropies observed, as we discuss below.  

\section{Cosmic ray density and large scale anisotropies}

We are interested in the first place in computing the dipolar component of the arrival direction distribution of cosmic rays coming from a source at a given distance $r_s$ in the presence of a turbulent magnetic field. In particular, we want to study its dependence on the energy of the particles  and consider the transition from  the diffusive propagation at low energies to the quasi-rectilinear propagation at high energies. 

In the diffusion regime, the density $n$ of ultra-relativistic particles propagating from a source located at ${\vec x}_s$ in an expanding universe obeys the equation
\begin{equation}
\frac{\partial n}{\partial t} + 3 H(t) n - b(E,t) \frac{\partial n}{\partial E} - n \frac{\partial b}{\partial E} - \frac{D(E,t)}{a^2(t)} \nabla^2 n = \frac{Q_s(E,t)}{a^3(t)} \delta^3({\vec x}-{\vec x}_s),
\label{difeq}
\end{equation}
where ${\vec x}$ denotes the comoving coordinates, $a(t)$ is the scale factor of the expanding universe, $H(t) \equiv {\dot a}/a$ is the Hubble constant and $D(E,t)$ is the diffusion coefficient. The  source function $Q_s(E,t)$ gives the number of particles produced per unit energy and time. The energy losses of the particles are described by 
\begin{equation}
\frac{{\rm d}E}{{\rm d}t}=-b(E,t),\ \ \ \ \ \ \ \ b(E,t)=H(t) E +b_{int}(E).
\label{eloss}
\end{equation}
This includes the energy redshift due to the expansion of the universe and energy losses due to the interaction with radiation backgrounds, that in the case of protons include pair production and photo-pion production due to interactions with the CMB background (see Appendix~II). 

In the static case (setting $H(t)=0$ and for time independent $D$) and in the absence of energy losses ($b = 0$) the solution for an impulsive source at $t=0$ ($Q_s(E,t)=Q_s(E) \delta(t)$) is given by
\begin{equation}
n(r_s,t,E)=\frac{Q_s(E) \exp(-r_s^2/(4Dt))}{(4\pi D t)^{3/2}},
\label{nstat}
\end{equation}
with $r_s=|\vec{x}-\vec{x}_s|$ being the comoving distance to the source. For a steady source the solution, obtained by integrating eq.~(\ref{nstat}) over time,  is given by  
\begin{equation}
n(r_s,E)=\frac{Q_s(E)}{4\pi r_s D(E)}.
\end{equation}
The solution for a static universe but including energy losses, $b_{int} \ne 0$,  was obtained by Syrovatsky \cite{si59}, and the general solution in an expanding universe was obtained by Berezinsky and Gazizov \cite{be07}, and is
\begin{equation}
n(E)=\int_0^{z_{max}}{\rm d}z\,\left|\frac{{\rm d}t}{{\rm d}z}\right| Q_s(E_g,z)\frac{\exp \left[-r_s^2/4\lambda^2\right]}{(4\pi\lambda^2)^{3/2}} \frac{{\rm d}E_g}{{\rm d}E},
\label{siro.eq}
\end{equation}
where $z_{max}$ is the maximum source redshift (note that in the diffusion regime redshift has the meaning of time rather than distance) and $E_g(E,z)$ is the original energy at redshift $z$ of a particle having energy $E$ at present ($z=0$). The source function  $Q_s$ will be assumed for definiteness to correspond to a power law spectra, $Q_s(E_g)=f(z) E_g^{-\gamma}$, up to a maximum energy $E_{max}$. In principle the source emissivity could depend on redshift through the factor $f(z)$, but for simplicity we will consider in this work non-evolving sources with constant $f(z)$. The Syrovatsky variable is given by
\begin{equation}
\lambda^2(E,z)=\int_0^{z}{\rm d}z\,\left|\frac{{\rm d}t}{{\rm d}z}\right| (1+z)^2\,D(E_g,z),
\end{equation}
with $\lambda(E,z)$ having the meaning of the typical distance diffused by CRs from the site of their production with energy $E_g(E,z)$ at redshift $z$ until they are degraded down to energy $E$ at the present time.
On the other hand, 
\begin{equation}
\left|\frac{{\rm d}t}{{\rm d}z}\right|=\frac{1}{H_0(1+z)\sqrt{(1+z)^3\Omega_m+\Omega_\Lambda}},
\end{equation}
where $H_0\simeq 70$~km/s/Mpc is the present Hubble constant, $\Omega_m\simeq 0.3$ the matter content and $\Omega_\Lambda\simeq 0.7$ the cosmological constant contribution at present. 

From the solution for the density just described, valid in the diffusive regime, the amplitude of the dipolar component of the arrival direction distribution can be obtained as 
\begin{equation}
{\vec \Delta} = \frac{3{\vec J}}{n}= 3 D \frac{{\vec \nabla} n}{n} = \Delta \ {\hat r}_s ,
\label{deldif}
\end{equation}
where ${\hat r}_s$ indicates that the dipole maximum points in the direction of the source. For the static case without energy losses and for a steady source, the dipole amplitude  reduces to
\begin{equation}
{ \Delta} = \frac{3 D(E)}{r_s}. 
\label{delst}
\end{equation}
When the effect of the universe expansion and/or the interaction energy losses are relevant, the dipolar anisotropy in the diffusive regime can be obtained from eq.~(\ref{deldif}) and directly differentiating the density in eq.~(\ref{siro.eq}). 

For very large energies the propagation becomes more rectilinear, so that the diffusion approximation ceases to be valid, with the arrival directions appearing increasingly clustered around the source location. In particular, in the limit of small deflections the dispersion of the arrival directions with respect to the source direction is given (in the static case and without energy losses) by \cite{ac99}
\begin{equation}
\langle \theta^2\rangle=\frac{(Ze)^2 B^2l_c r_s}{6 E^2}=\frac{r_s}{6 l_c}\left(\frac{E_c}{E}\right)^2.
\label{rmsdef}
\end{equation}

In the general case the dipolar component of the anisotropy can be computed as follows. The distribution of the arrival directions $\hat u$ of particles from a source at ${\vec r}_s\equiv r_s\hat{r}_s$ only depends on the angle between $\hat u$ and ${\vec r}_s$, $\theta={\rm acos}(\hat u\cdot{\hat r}_s)$, and can be expanded in Legendre polynomials as
\begin{equation}
\Phi(\hat u)= f(\cos \theta)=\Phi_0 + \Phi_1 \hat u\cdot{\hat r}_s + \dots.
\label{fluxexp} 
\end{equation}
The expansion coefficients can be computed from
\begin{eqnarray}
\Phi_0&=&\frac{1}{4\pi}\int \Phi(\hat u) {\rm d} \Omega = \frac{1}{2}\int_{-1}^1f(\cos \theta){\rm  d}\cos\theta,\nonumber\\
\Phi_1&=&\frac{3}{4\pi}\int \Phi(\hat u) \hat u\cdot \hat{r}_s {\rm d} \Omega = \frac{3}{2}\int_{-1}^1f(\cos \theta) \cos \theta {\rm d}\cos\theta.
\label{coexp}
\end{eqnarray}
The dipole  amplitude is then given by
\begin{equation}
\Delta=\frac{\Phi_1}{\Phi_0} = 3 \langle \cos \theta \rangle
\label{dipamp}
\end{equation}
and points towards the source direction. For perfectly rectilinear propagation one has
$\langle \cos \theta \rangle =1$ so that $\Delta =3$. This is the expected value as the distribution is a delta function in the source direction and its expansion in Legendre polynomials would correspond to
\begin{equation}
\delta(\Omega) = \frac{1}{4 \pi} \left(P_0(\cos \theta) + 3 P_1(\cos \theta) +\dots\right) = \frac{1}{4 \pi} (1 + 3 \cos \theta +\dots ).
\end{equation}
In the quasi rectilinear regime of small deflections, replacing $\langle \cos \theta \rangle \simeq 1 - \langle \theta^2 \rangle /2 $
and using eq.~(\ref{rmsdef}), the dipolar anisotropy can be written as
\begin{equation}
\Delta \simeq 3\left(1-\frac{r_s}{12 l_c}\left(\frac{E_c}{E}\right)^2\right).
\label{DeltaQR}
\end{equation}

In the next section we will compute the dipolar anisotropy in the full range of energies and distances to the source, covering the transition from spatial diffusion to the quasi-rectilinear regime using numerical simulations of particles propagating in a turbulent magnetic field. By following many particles of a given energy in a turbulent magnetic field $\langle \cos \theta \rangle$ can be computed as the mean cosine of the angle between the original direction of the CR velocity and the vector describing its position  when the particles pass at a distance $r_s$ from the original point. In this way the dipolar anisotropy can be numerically obtained for all energies, and we will be able to match the results from the diffusive and quasi-rectilinear regimes.

\section{Simulations of charged particle propagation in a turbulent magnetic field}

The evolution of the direction of propagation $\hat n$ of particles with charge $Ze$ in the turbulent field is followed by integrating the Lorentz equation 
\begin{equation}
\frac{{\rm d}{\hat n}}{{\rm d}t}=\frac{Z e c}{E(t)} {\hat n} \times {\vec B} ({\vec x, t}).
\label{lorentz}
\end{equation}
We will only consider the case of protons in the following, thus $Z=1$, although if energy losses can be ignored all results also apply to the case of nuclei by replacing $E$ by $E/Z$ (the inclusion of energy losses in the case of nuclei is complicated by the fact that photo-disintegration processes change the nuclear masses and lead to the production of secondary nucleons).
 The presence of the magnetic field does not change the magnitude of the velocity (nor the particle energy), it only modifies the propagation direction. The dependence with time appearing in eq.~(\ref{lorentz}) arises due to the redshift in the expanding universe and from energy losses due to the interaction of the protons with the CMB radiation. We also included in eq.~(\ref{lorentz}) a possible evolution of ${\vec B}$ with time.

We will first consider the static case, neglecting the time variation of both the energy  $E$ and the magnetic field ${\vec B}$. We solve the propagation equation for a large number of particles using two different approaches, a full numerical integration of eq.~(\ref{lorentz}) in particular realizations of the turbulent magnetic field and an integration of a stochastic differential equation describing the scattering of charged particles in randomly oriented magnetic cells.

\subsection{Full numerical integration}

\begin{figure}[t]
\centerline{\epsfig{width=3.5in,angle=-90,file=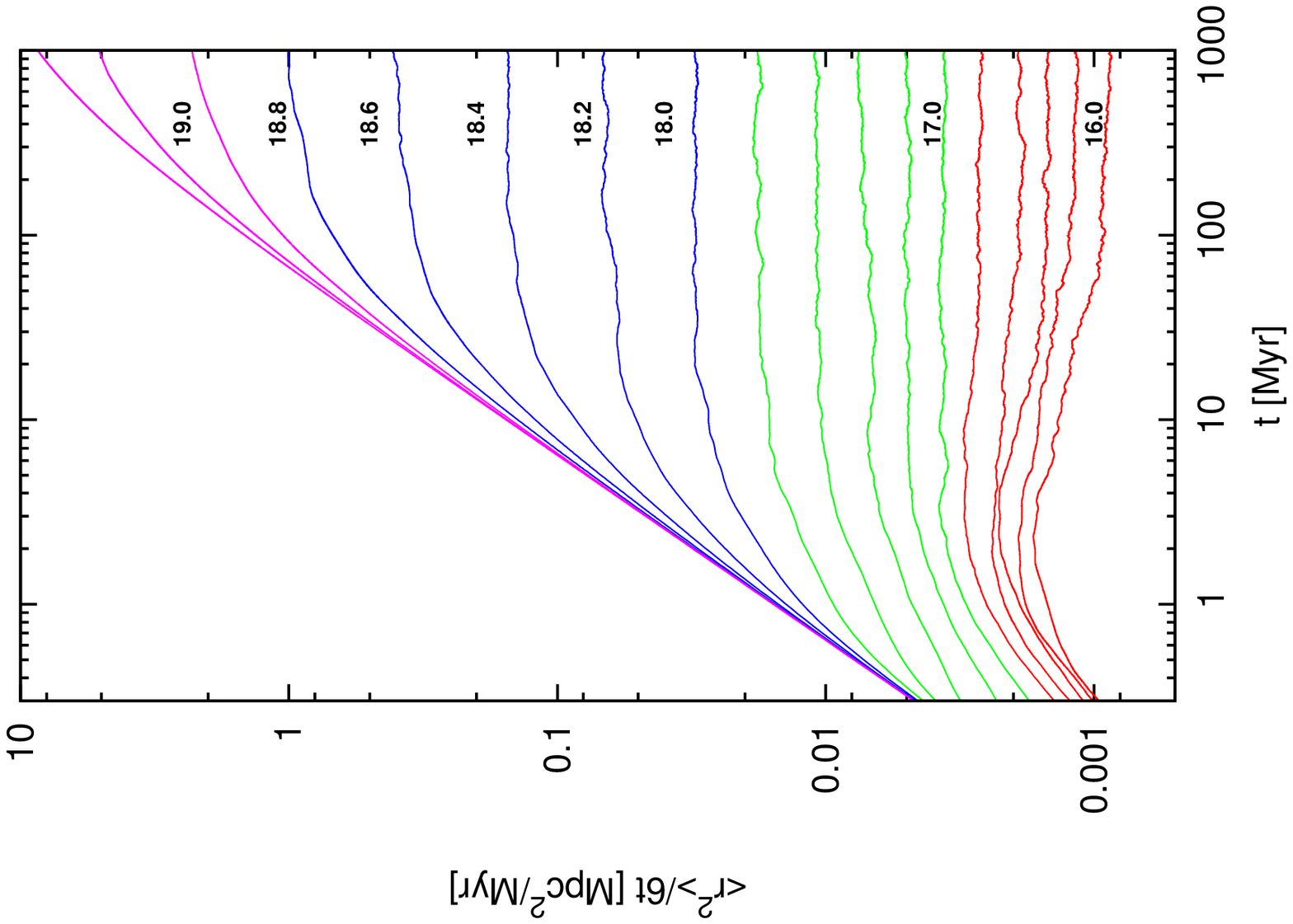}\epsfig{width=3.5in,angle=-90,file=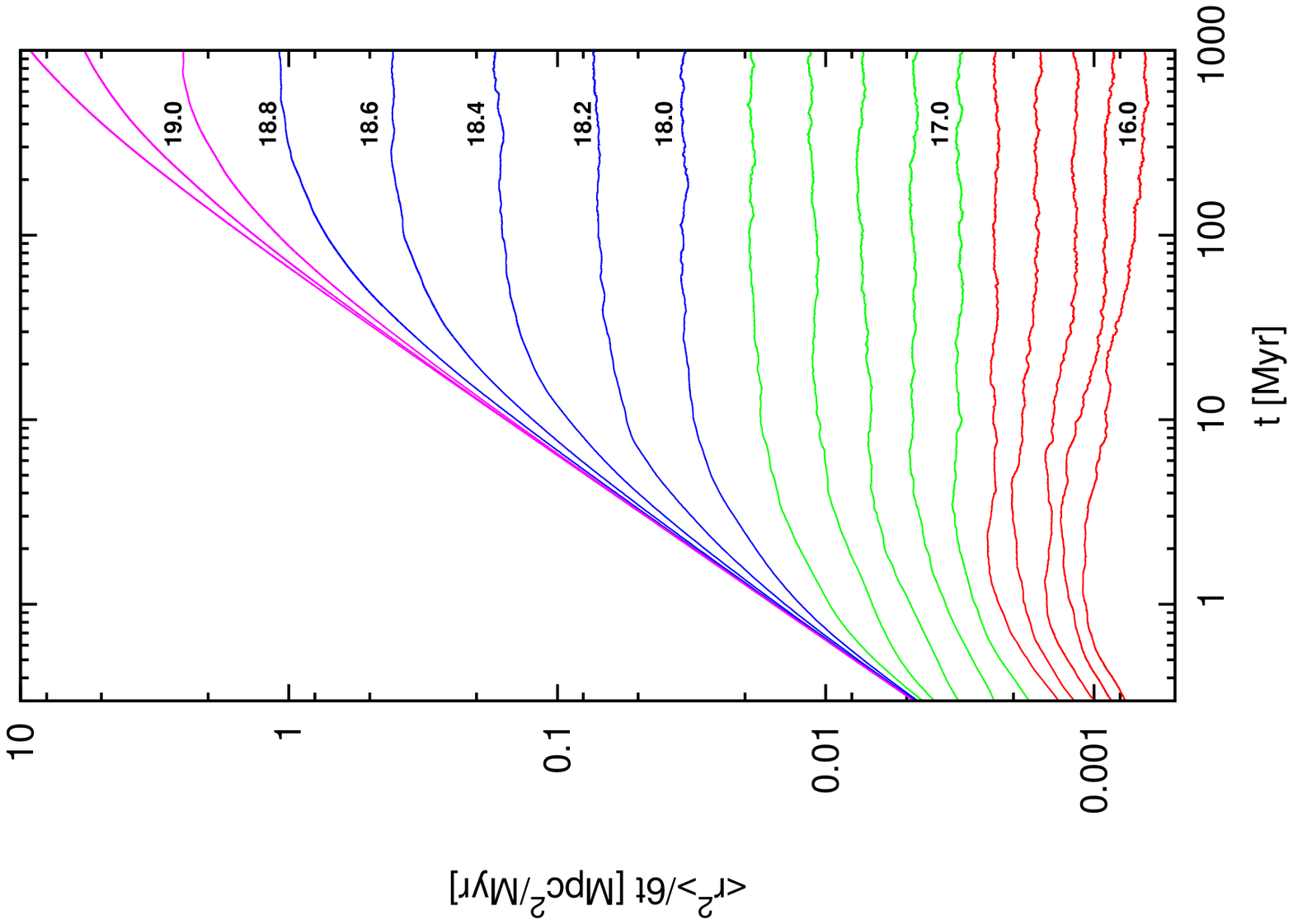}}
\vskip 1.0 truecm
\caption{Values of $ \langle r^2(t) \rangle /6t$ as a function of time evaluated from numerical integration of the Lorentz equation for several values of the energy of protons around $E_c$. The turbulent magnetic field has strength $B=10$~nG and maximum scale of turbulence $L_{max}=1$~Mpc. The magnetic energy distributions are of the  Kolmogorov (left panel) and Kraichnan (right panel) type. The diffusion coefficient $D(E)$ is the value at the plateau. The logarithm of the energy (in units of eV) is displayed over some of the plateaus. } 
\label{r2t}
\end{figure}

\begin{figure}[t]
\centerline{\epsfig{width=1.75in,angle=-90,file=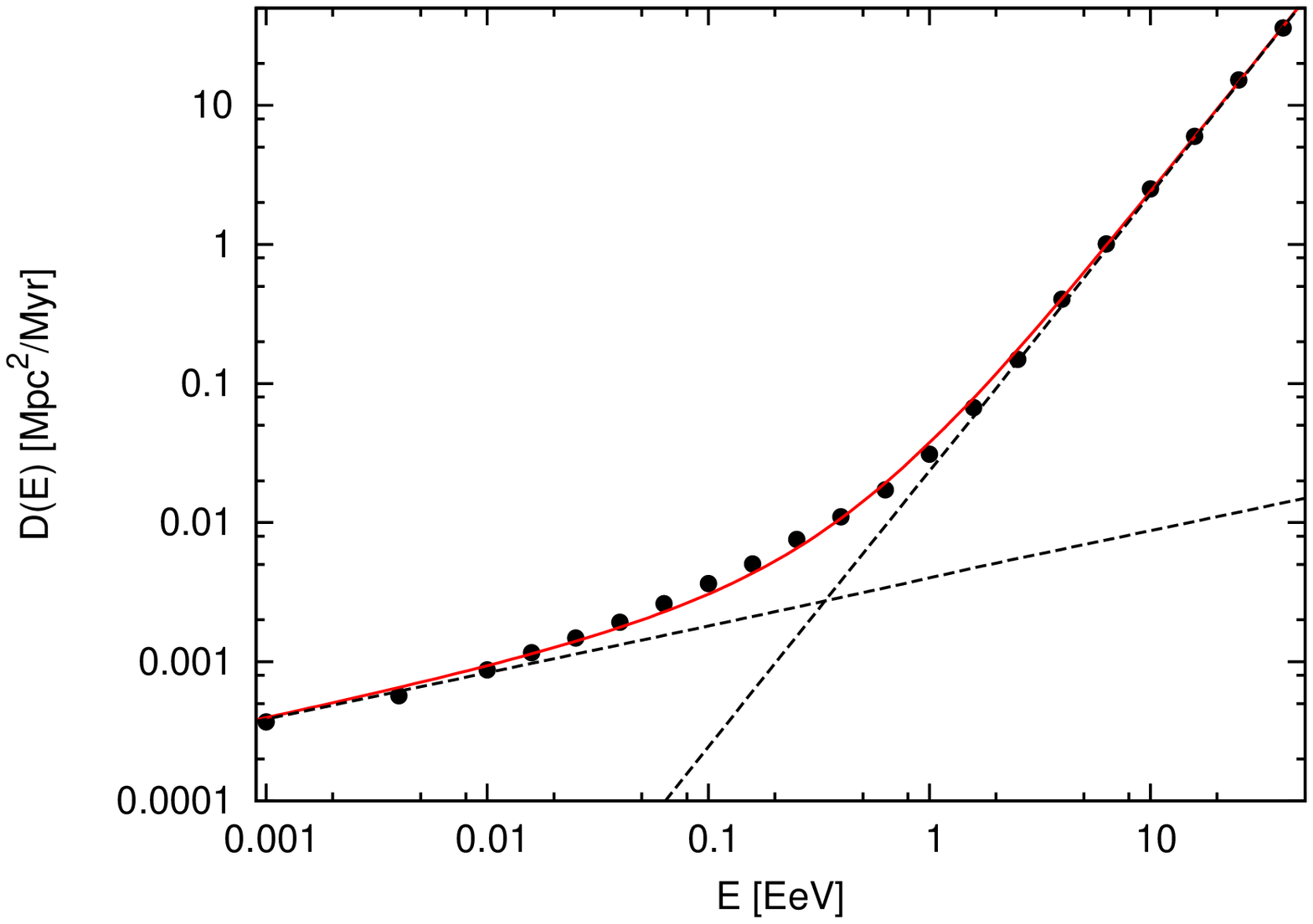}\epsfig{width=1.75in,angle=-90,file=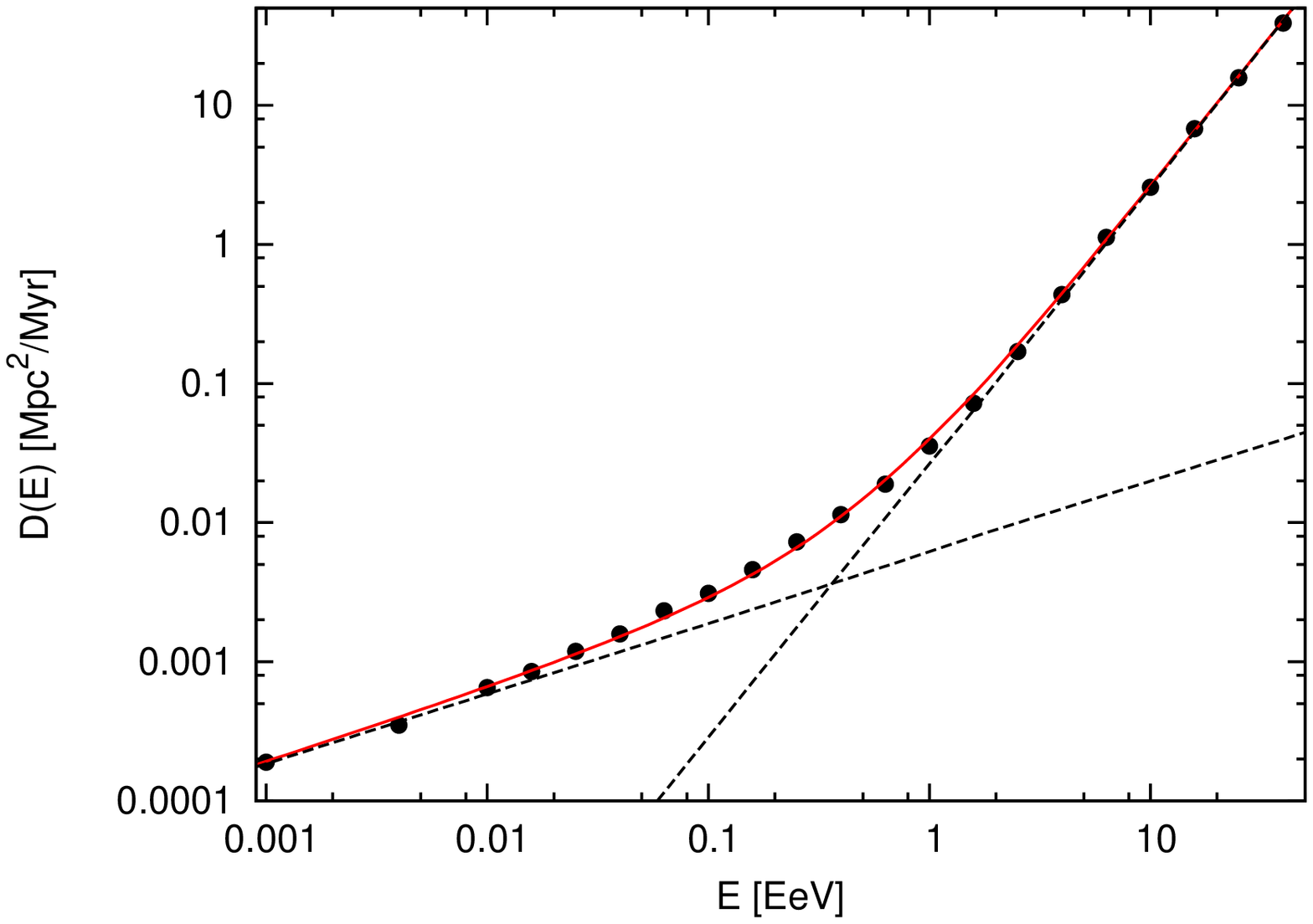}}
\vskip 1.0 truecm
\caption{Diffusion coefficient $D(E)$  (in units of ${\rm Mpc}^2/{\rm Myr}$) evaluated numerically (dots) and fit through eq.~(\ref{D(E).eq}) (solid line). The field parameters are as in Fig. \ref{r2t}. The asymptotic values $\frac{4c}{3}l_c\left({E}/{E_c}\right)^2$ at high energies and $a_L\frac{c}{3}l_c\left({E}/{E_c}\right)^{(2-m)}$ at low energies are also shown (dotted lines). The left panel corresponds to a Kolmogorov spectrum ($m=5/3$, $a_L\approx 0.23$) and the right panel to a Kraichnan energy distribution ($m=3/2$, $a_L\approx 0.42$).}
\label{D(E).fig}
\end{figure}

In our first approach, the trajectories of charged particles are solved by numerical integration of the Lorentz equation~(\ref{lorentz}). The turbulent magnetic field is modeled by a superposition of Fourier modes as described in Appendix I. No energy losses are included in this case. We follow the trajectories of a large number of particles, each in a different realization of the turbulent magnetic field. At each integration step we evaluate the rectilinear distance $r$ between the origin and the particle position, as well as the angle $\theta$ between its initial velocity and its position vector. We show in Figure~\ref{r2t} the results for the quantity  $\langle r^2(t)\rangle/6t$, averaged over all the simulated trajectories, in units of ${\rm Mpc}^2/{\rm Myr}$, as a function of the elapsed time $t$ and for several values of the energy around $E_c$. The field parameters were chosen for these simulations as in ref.~\cite{gl07}: $B= 10$ nG, $L_{\rm max}= 1$ Mpc. The left panel corresponds to a Kolmogorov spectrum of turbulence while the right panel to a Kraichnan spectrum.  The results correspond to protons, but also apply to nuclei with $E$ replaced by $E/Z$ (since energy losses are not considered here).
The diffusion coefficient $D(E)$ is the value of $\langle r^2 \rangle /6t$ at the plateau, when sufficient time has elapsed for the particles to reach the diffusive regime. The values of $D$ as a function of the energy are shown in Figure~\ref{D(E).fig}. 
The results for the case of Kolmogorov turbulence are comparable to those in ref.~\cite{gl07}. As anticipated, $D\propto E^2$ for $E > E_c$ and $D\propto E^{2-m}$ at sufficiently low energies. We will show in the next section that a stochastic approach allows an analytic derivation of the diffusion coefficient at high energies, given by $D(E) = \frac{4}{3}cl_c\left({E}/{E_c}\right)^2$, which is verified by the numerical solutions. We then fit the numerical results with the function
\begin{equation}
D(E) = \frac{c}{3}l_c\left[ 4\left(\frac{E}{E_c}\right)^2 + a_I\left(\frac{E}{E_c}\right) + a_L\left(\frac{E}{E_c}\right)^{2-m}\right]
\label{D(E).eq}
\end{equation}
A term with a linear dependence upon the energy, interpolating the transition between the resonant and non-resonant diffusion regimes, was added to improve the fit around $E_c$.\footnote{The physical origin of the interpolating term proportional to $a_I$ might be related to the residual effects on the resonant diffusion of the $B$-modes with long wavelength, which may generate non-isotropic diffusion and also lead to drift effects (the Hall diffusion scaling indeed linearly with $E$).}  The coefficients $a_L$ and $a_I$ are obtained from a fit to the numerical results. For a Kolmogorov spectrum ($m=5/3$) $a_L\approx  0.23$ and $a_I\approx  0.9$. In the case of a Kraichnan spectrum ($m=3/2$), $a_L\approx  0.42$ and $a_I\approx  0.65$
%(G.A.P.: a_L=(2 pi)**(1-m)= .29 Kolmogorov and .4 Kraichnan)

\begin{figure}[t]
\centerline{\epsfig{width=2.5in,angle=-90,file=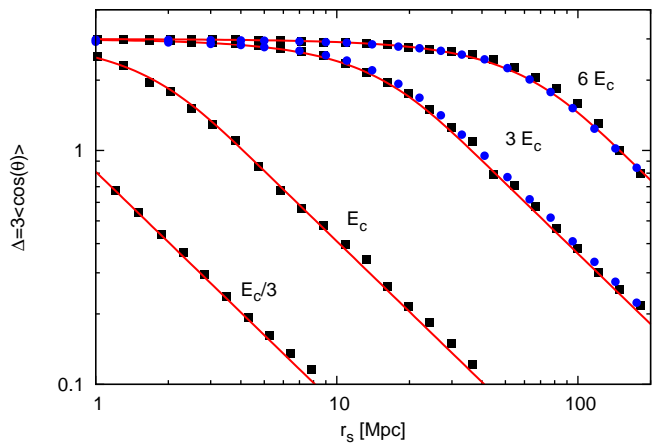}}
\vskip 1.0 truecm
\caption{Anisotropy $\Delta=3\langle \cos \theta \rangle$  as a function of the distance to the source $r_s$, for some representative values of the energy: $E/E_c= 1/3,1,3,6$. The fit provided by eq.~(\ref{delfit}) is shown by the solid line. The result obtained through numerical solution of the trajectories of charged particles in the turbulent field is shown by squares. A Kolmogorov spectrum was used for illustration, with $B=3$~nG and $l_c= 1$~Mpc. The critical energy is $E_c\approx 2.7$~EeV. The result from the numerical integration of the stochastic approach is shown by dots (for $E/E_c= 3,6$). } 
\label{Delta(E).fig}
\end{figure}

The anisotropy in the distribution of arrival directions as a function of the distance to the source and the energy of the particles was also evaluated numerically. Each time a trajectory crossed a surface of radius $r_s$ centered at the source we evaluated $\cos\theta$, and then we averaged over all the trajectories and over all the crossings for each trajectory.  We verified that, as expected, $\Delta=3\langle \cos \theta \rangle=3D(E)/r_s$ for $r_s \gg l_D$, when the spatial diffusion regime is reached. At large distances the dipole amplitude $\Delta$ also depends on the time of source activity, and is larger than its asymptotic value if the stationary regime has not yet been reached. This is related to the ``magnetic horizon'' effect \cite{le04,be06,gl07,mo13}, that prevents particles  reaching beyond a certain distance even if the source was active over Hubble times. This dependence is not relevant for our present purposes, since these situations also require consideration of energy losses, which we will do through eq.~(\ref{siro.eq}). Here we are interested in the transition from quasi-rectilinear to diffusive propagation. We illustrate in Figure~\ref{Delta(E).fig} the results for the anisotropy $\Delta$ as a function of the distance to the source for several representative values of the energy. 
Along with the numerical results we plot a function that provides a good fit of $\Delta$ as a function of $E$ and $r_s$:     
\begin{equation}
\Delta(E,r_s) \simeq \frac{3 D(E)}{cr_s}\left[1-\exp\left(-\frac{cr_s}{D(E)}-\frac{7}{18}\left(\frac{cr_s}{D(E)}\right)^2\right)\right] \ .
\label{delfit}
\end{equation}
For large values of $r_s$ it tends to the value in eq.~(\ref{delst}) valid for diffusion, while the expansion to first-order in $cr_s/D(E)$ coincides with the approximation in eq.~(\ref{DeltaQR}) valid for the quasi-rectilinear regime of small deflections over a coherence length. The complete function provides a good fit across the different regimes, from the quasi-rectilinear propagation to the full spatial diffusion.

\subsection{Stochastic differential equation}

The propagation of protons in the presence of a turbulent homogeneous and isotropic field can also be simulated by numerically integrating a stochastic differential equation as proposed in ref.~\cite{ac99}. Within this approach it is possible to describe the propagation in the non-resonant regime when the deflection in a distance equal to the coherence length $l_c$ is small, including both the quasi-rectilinear propagation regime and a regime of spatial diffusion if the  propagation distance is larger than the diffusion length.

The magnetic scatterings of the protons lead to angular diffusion of the propagation direction $\hat n$ with an angular diffusion coefficient given by \cite{ac99}
\begin{equation}
{\cal D} _{ij} \equiv \frac{\langle \Delta n_i \Delta n_j\rangle}{2 c \Delta t}
=\frac{l_c}{8} \left(\frac{Ze B}{E}\right)^2 P_{ij} \equiv {\cal D}_0 P_{ij},
\label{dij}
\end{equation}
where $P_{ij} \equiv (\delta_{ij} - n_i n_j)$ is the tensor projecting to the plane orthogonal to $\hat n\equiv (n_1,n_2,n_3)$.

The evolution of the unit vector $\hat n$ is followed by numerically integrating a stochastic differential equation. At each step the particle moves a distance $l_c$ and its direction of propagation suffers a stochastic change 
\begin{equation}
(\Delta \hat n)_i = \sqrt{2 l_c {\cal D}_0} P_{ij} \xi_j,
\label{dni}
\end{equation}
where repeated indices are summed and ($\xi_1, \xi_2, \xi_3$) are three Wiener processes drawn at each step from a Gaussian distribution with unit dispersion, so that $\langle \xi_j \rangle =0$ and  $\langle \xi_j^2 \rangle =1$. 

The new propagation direction is obtained as
\begin{equation}
\hat n (t+\Delta t) = \sqrt{1-|\Delta \hat n|^2} \ \hat n (t) + \Delta \hat n,
\label{nstoch}
\end{equation}
which preserves the norm of $\hat n$. The new position of the particle is given by
\begin{equation}
{\vec x} (t+\Delta t) = {\vec x} (t) + l_c \hat n (t).
\label{xstoch}
\end{equation}
\begin{figure}[t]
\centerline{\epsfig{width=3in,angle=-90,file=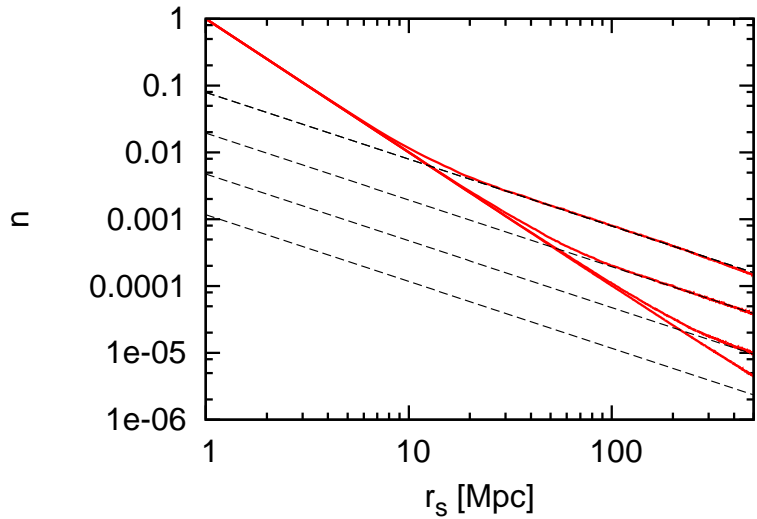}}
\vskip 1.0 truecm
\caption{Cosmic ray density as a function of the distance to the source for 
values of the ratio between the energy and the critical energy of $E/E_c = 3,6,12,24$ (solid lines, from top to bottom). Dashed lines show the $r^{-1}$ dependence valid  in the regime of spatial diffusion.} 
\label{nr}
\end{figure}
By following a large number of particles for a large enough time and counting the number of times that the particles cross spherical caps of different radius $r$ around the initial point, the density of particles as a function of the distance to the source can be determined. This is illustrated in Figure \ref{nr} for different values of the ratio $E/E_c$, showing the transition from $n(r) \propto r^{-2}$ at small $r$, corresponding to the quasi-rectilinear propagation, to $n(r) \propto r^{-1}$ in the diffusive regime. The transition between these regimes takes place at a radius $r_t$ which becomes larger for increasing energies: $r_t\simeq l_c (E/E_c)^2$. The value of the diffusion coefficient can be obtained from the plateau value in the $D \simeq \langle r^2(t)\rangle/6 t$ vs. $t$ plot, as described in the previous section. The results of the stochastic simulations nicely match the full propagation results in the regime in which the deflections are small in each coherence length distance. In Appendix B of ref. \cite{ac99} an analytic solution was obtained as
\begin{equation}
\langle r^2(t) \rangle = \frac{1}{{\cal D}_0} \left[c t - \frac{1}{2{\cal D}_0}\left(1 - \exp(-2 {\cal D}_0 c t)\right)\right], 
\label{r2}
\end{equation}
with ${\cal D}_0$ defined in eq.~(\ref{dij}) and given by
\begin{equation}
{\cal D}_0=\frac{1}{8 l_c} \left(\frac{E_c}{E}\right)^2.
\end{equation}
The spatial diffusion coefficient is obtained from the limiting value  of  $\langle r^2(t)\rangle/6 t$ for $ct \gg {\cal D}_0^{-1}$, and using  eq.~(\ref{r2}) one finds that
\begin{equation}
D(E) = \frac{c}{6 {\cal D}_0} = \frac{4c}{3}\left(\frac{E}{E_c}\right)^2 l_c,
\end{equation}
corresponding to the high energy limit in eq.~(\ref{D(E).eq}).  This implies in particular that the coefficient $a_H$ introduced in Section~2 is $a_H=4$.

Finally, by computing the mean cosine of the angle between the original direction and the position vector of the particles when they cross a cap at a given radial distance $r$ one can obtain the dipolar anisotropy from eq.~(\ref{dipamp}). This is shown in Figure~\ref{Delta(E).fig}, which also displays  the results from the full numerical simulation as well as the approximation in eq.~(\ref{delfit}), and a very good agreement is indeed obtained.

Within the framework of these stochastic simulations it is not difficult to take into account the effects of the expansion of the universe and the proton energy losses due to interactions with the CMB radiation. The coordinate $\vec r$ now refers to the comoving coordinate. In the case of the expanding universe we can include the possible variation of the magnetic field amplitude with the redshift, that we can take as $B(z) = B(0) (1+z)^{2-\mu}$, where the factor $(1+z)^2$ arises from the flux conservation and the index $\mu$ was introduced in ref.~\cite{be07} to account for magneto-hydrodynamic effects, and was taken there as $\mu=1$, as we will adopt in the simulations. The critical energy dependence with $z$ is $E_c(z)=E_c(0) (1+z)^{1-\mu}$. The coherence length will typically scale as $l_c(z)=l_c(0)/(1+z)$. 

We backtrack protons from the observation time at $z =0$ in time steps adapted such that the change in the comoving coordinates $|\Delta {\vec r}|$ is equal to $l_c(0)$ in each step, $c|\Delta t| = |\Delta {\vec r}|/(1+z)$. This corresponds to a step in redshift $|\Delta z| = l_c(0) \sqrt{\Omega_m (1+z)^3+\Omega_\Lambda} H_0/c$. Due to energy losses from redshift and from the interactions with the CMB we have to take into account that particles arriving to $z=0$ with energy $E(0)$ had a larger energy $E_g(z)$ at each previous step, and the inclusion of these effects is performed as described in Appendix II. Then, in each step the particle moves a comoving distance $l_c(0)$ and the propagation direction suffers a stochastic change given by
\begin{equation}
(\Delta \hat n)_i = \sqrt{2 {\cal D}_0(z) l_c(0)/(1+z)} P_{ij} \xi_j,
\label{dnie}
\end{equation}
where
\begin{equation}
{\cal D}_0(z)=\frac{1}{8 l_c(0)} \left(\frac{E_c(0)}{E_g(z)}(1+z)^{3-2\mu}\right)^2.
\end{equation}
The new propagation direction is obtained from 
\begin{equation}
\hat n (z+\Delta z) = \sqrt{1-|\Delta \hat n|^2}\ \hat n (z) + \Delta \hat n
\label{nstoch2}
\end{equation}
and the new comoving coordinate from
\begin{equation}
{\vec x}(z+\Delta z) = {\vec x}(z) +l_c(0) {\hat n} (z).
\end{equation}
In this way we can follow the proton trajectories back in time. In order to compute the expected dipole anisotropy of particles from a source at comoving distance $r_s$,  for any given arrival energy $E(0)$ we backtrack the trajectories of a large number of particles and compute the mean cosine of the angle between the initial direction and the position when the particles pass at a comoving distance $r_s$ from the original point. When taking the mean we have to include a weight factor for each particle equal to $[E_g(z)/E(0)]^{-\gamma} {\rm d}E_g/{\rm d}E$, where the first factor takes into account that for a source emitting protons with a spectrum $E^{-\gamma}$ there will be less particles with the higher energy $E_g(z)$ required to reach the observer with a given $E(0)$, and the second factor takes into account the change in the energy bin width from the emission to the observation.

\begin{figure}[t]
\centerline{\epsfig{width=2.2in,angle=270,file=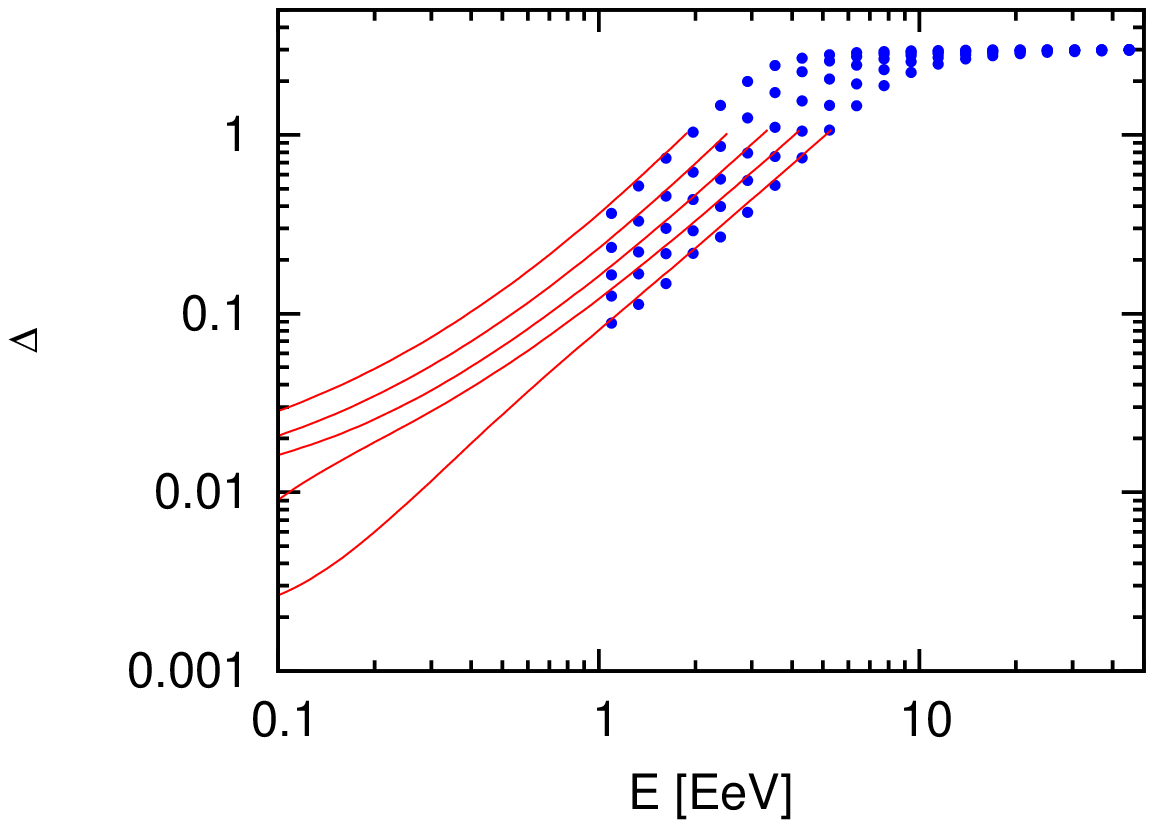},\epsfig{width=2.2in,angle=270,file=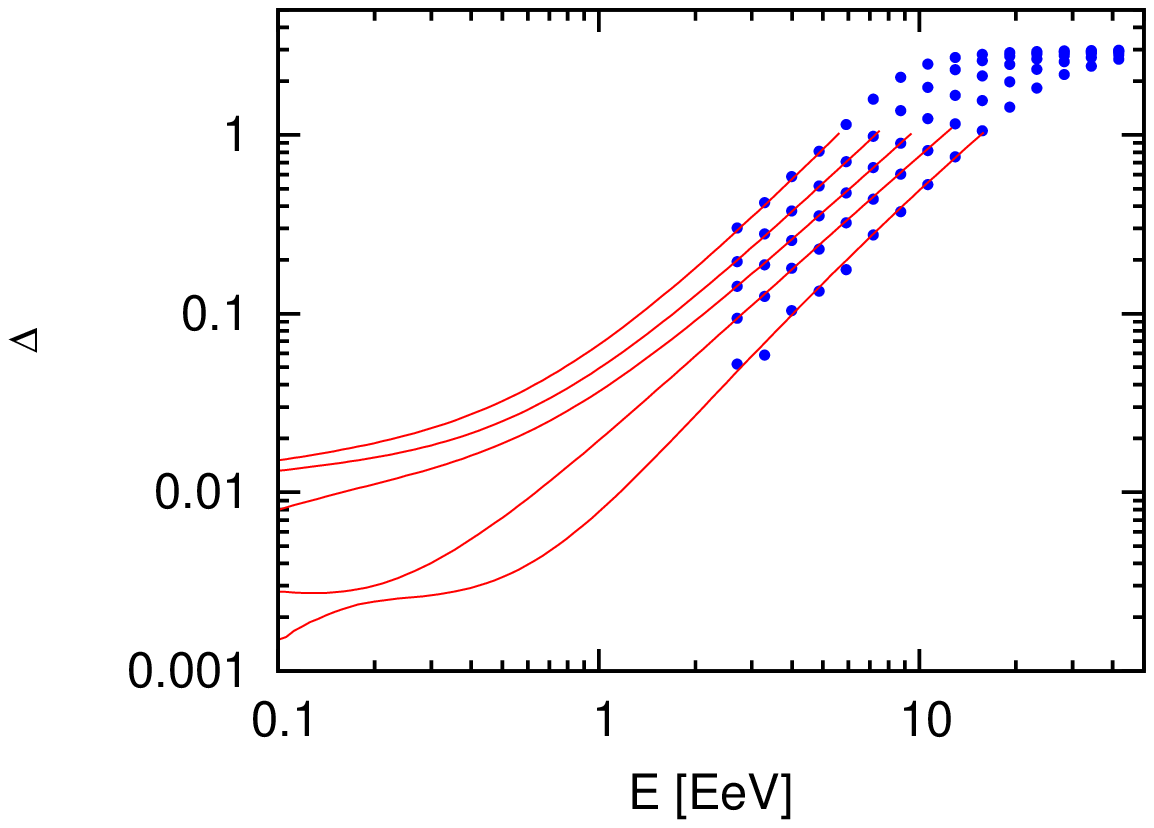}}
\vskip 1.0 truecm
\caption{Dipole amplitude for one source located at a comoving distance of 25, 50, 100, 200 and 400 Mpc (from top to bottom) for a coherence length of the magnetic field of $l_c= 1$ Mpc and an amplitude $B=1$ nG (left panel) and $B=3$ nG (right panel).} 
\label{dsedep}
\end{figure}

We show in Figure~\ref{dsedep} the dipole anisotropy resulting from one single source located at a comoving distance of 25, 50, 100, 200 or 400 Mpc (from top to bottom) as a function of the energy. Blue dots show the results obtained from the integration of the stochastic differential equation while solid lines show the results from the solution to the diffusion equation (using eqs.~(11) and (8)). A very good agreement is seen in the overlapping region. A spectral index $\gamma=2$ and a maximum energy $E_{max}=10^{21}$~eV are considered in all the examples. The magnetic field coherence length was taken as $l_c= 1$ Mpc. In the left panel we considered $B$= 1 nG, leading to $E_c=0.9$ EeV, while in the right panel we considered $B$= 3 nG, corresponding to $E_c = 2.7$ EeV. A close similarity of the results in both plots is apparent when the energy is rescaled by a factor 3, i.e. corresponding to the same values of $E/E_c$, although due to the dependence of the energy losses with energy the scaling is not exact.

\section{Large scale anisotropy from many sources}

In the previous section the dipolar anisotropy from an individual source in the presence of a turbulent magnetic field was computed as a function of the source distance and of the CR energy. In a realistic situation the total cosmic ray flux will probably originate from a set of several (or many) sources. The total dipolar component of the flux will mainly depend on the location and intensities of the nearest sources and on whether there is an inhomogeneous distribution of the sources at large scales. If there are several sources contributing to the flux, the dipolar anisotropy can be obtained from the superposition of the individual source dipoles  through
\begin{equation}
{\vec \Delta} (E)=\sum_{i=1}^N \frac{n_i}{n_t}(E) {\vec \Delta}_i (E),
\label{diptot}
\end{equation}
where $N$ is the number of sources giving a non-negligible contribution to the flux at energy $E$, $n_i/n_t$ measures the fraction of the flux coming from the $i$-th source and ${\vec \Delta}_i (E)$ is the dipole anisotropy of the flux from source $i$ computed in the previous section and shown in Figure \ref{dsedep}.

\begin{figure}[t]
\centerline{\epsfig{width=2.2in,angle=270,file=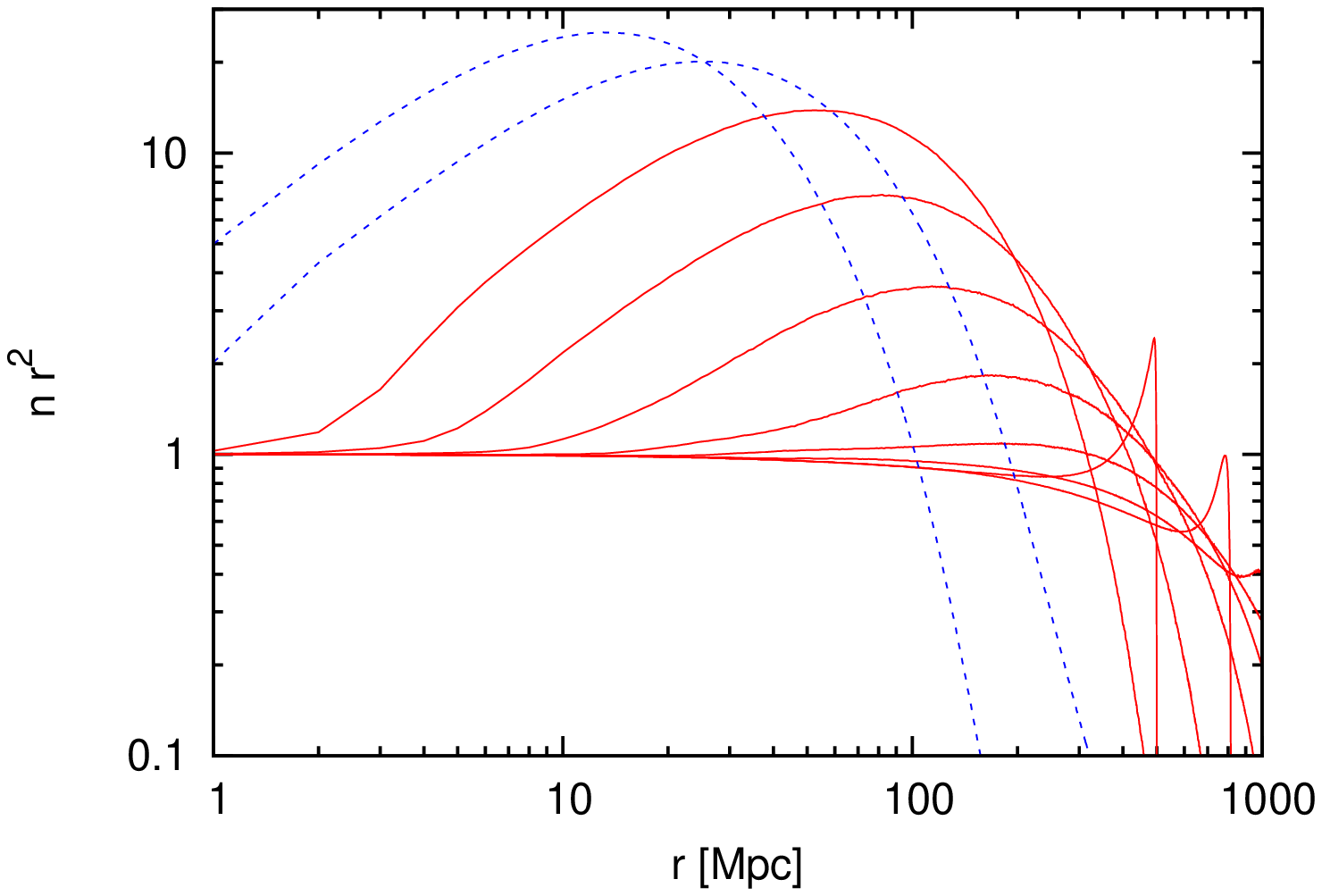},\epsfig{width=2.2in,angle=270,file=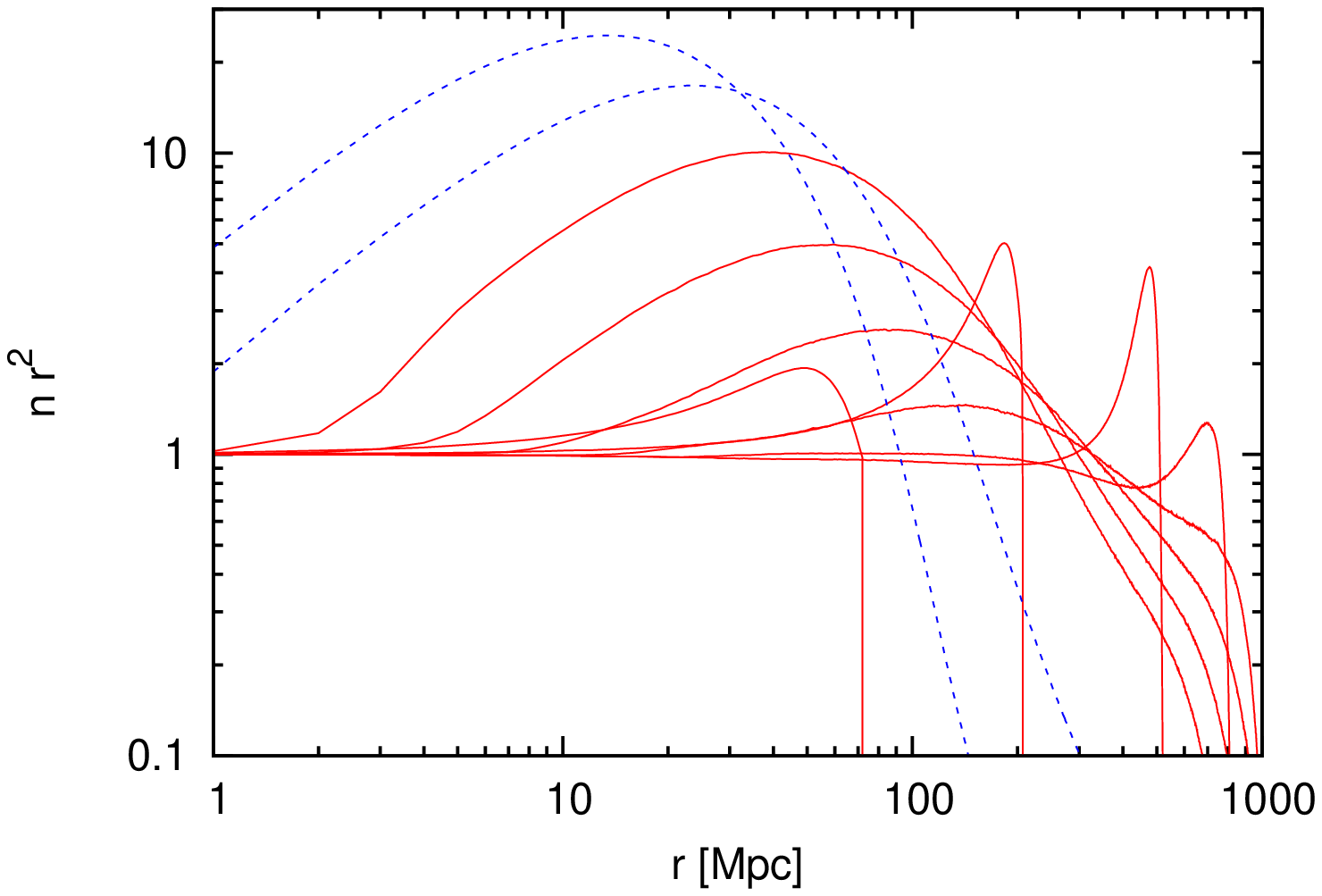}}
\vskip 1.0 truecm
\caption{Density of cosmic rays $n(r)$ times $r^2$ as a function of the distance from the source to the observer for different observed energies. Left panel: for $B=1$ nG and $E = 0.9, 1.5, 2.6, 4.4, 7.4, 12.5, 21.2$ and 36 EeV (red solid lines from top-left to bottom-right) and $E = 0.1$ and 0.3 EeV (blue dashed lines). Right panel: same for $B=3$ nG and $E = 2.7, 4.6, 7.7, 13.1, 22.2, 37.6, 63.8$ and 108 EeV (red solid lines) and $E=0.3$ and 0.95 EeV (blue solid lines).}
\label{nr2}
\end{figure}

In order to estimate the relative contribution to the anisotropy of the different sources as a function of their distance to the observer we will make the simplifying assumption that the sources are steady and have equal intrinsic intensities, so that for each energy the relative contribution to the flux from different sources  will only depend on the distance to the source $r_i$. The product $n(r) r^2$ is shown in Figure \ref{nr2} for different energies (arbitrary normalization).  
At low energies, where particles are diffusing, we can see a large enhancement of the flux with respect to the typical $n \propto r^{-2}$ behavior characteristic of rectilinear propagation, followed by a drop at large distances corresponding to the magnetic horizon effect. For large energies the diffusion enhancement disappears as particles travel more straight and they can also arrive from larger distances. For the largest energies the maximum distance from which sources contribute is limited  due to energy losses (GZK horizon), and a bump in the flux appears at large distances  due to the pile-up caused by the effect of the photo-pion production threshold. We note that even if we assumed the magnetic field turbulence to be uniform everywhere, which may be a crude approximation at very large distances due to the presence of voids and filamentary structures in the matter distribution, at the  energies for which distances much larger than $\sim 100$~Mpc are relevant, the overall contribution from far away sources is not expected to be strongly affected by diffusion effects.

The fact that the sources are distributed in different sky directions means that the vector sum in eq.~(\ref{diptot}) will generally lead to a smaller dipole amplitude when many sources contribute. In the case that only few sources are relevant, the direction of these particular sources will determine the dipolar anisotropy, while if many sources are relevant, the overall large scale distribution of the sources, in particular whether the distribution has a non-vanishing dipole component, can have a significant effect. 
\begin{figure}[t]
\centerline{\epsfig{width=2.2in,angle=270,file=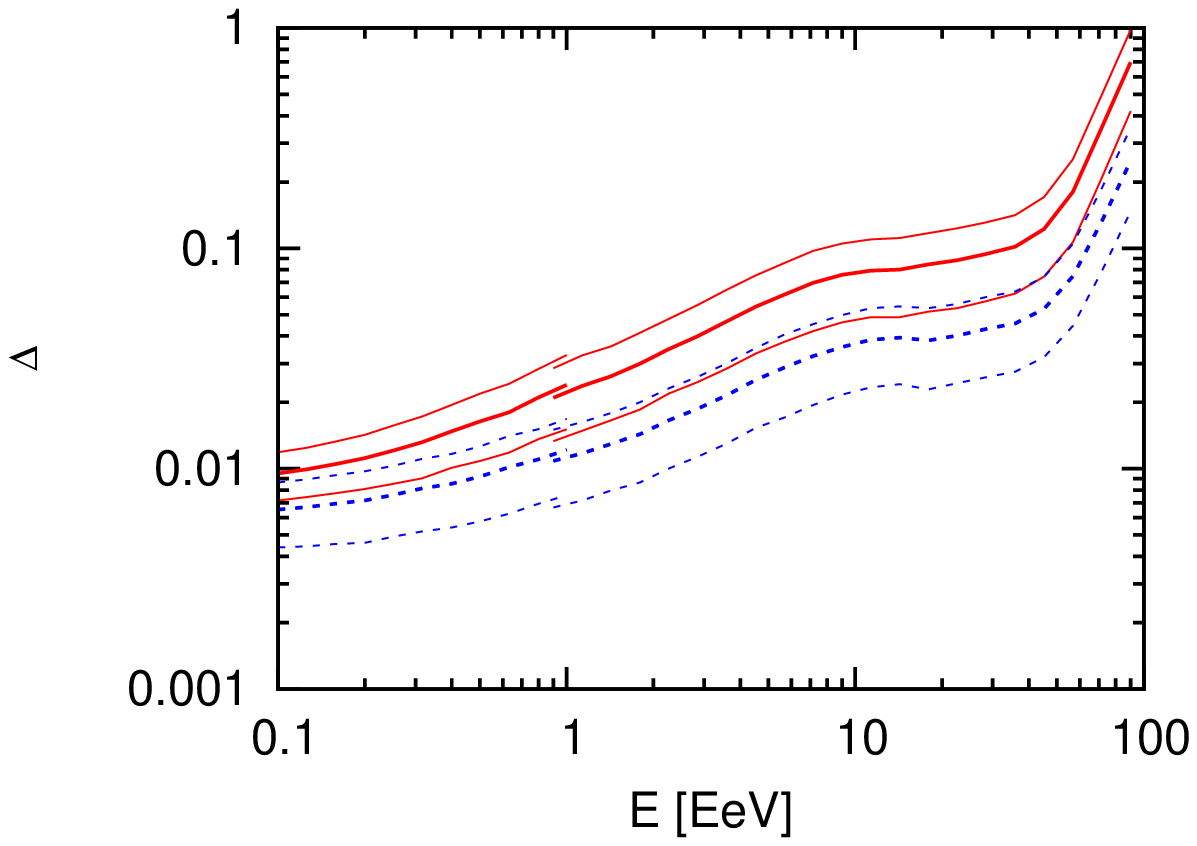},\epsfig{width=2.2in,angle=270,file=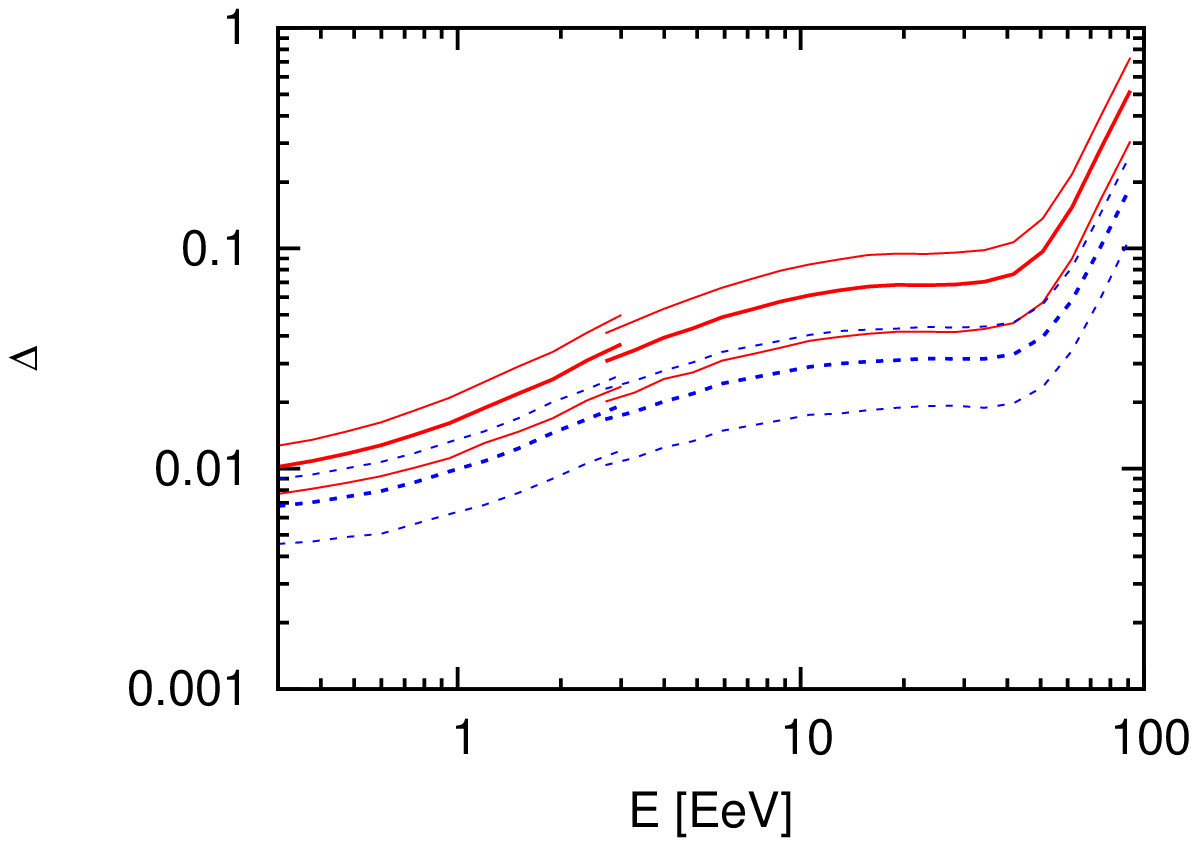}}
\vskip 1.0 truecm
\caption{Mean and dispersion of the total dipole amplitude as a function of the energy. Left panel: for a turbulent magnetic field of $B= 1$ nG and a density of sources $\rho = 10^{-5}$ Mpc$^{-3}$ (red solid lines) and $\rho = 10^{-4}$ Mpc$^{-3}$ (blue dashed lines). Right panel: same for $B= 3$ nG. }
\label{delt}
\end{figure}

In order to quantify the total amplitude of the dipolar anisotropy we performed some simple simulations. Starting with one source at a random direction in the sky, that represents the closest source, we subsequently added new sources in random directions and computed the new total dipolar anisotropy using eq.~(\ref{diptot}). The radial distances from the observer to the sources are taken as the mean expected value for the $i$-th closest source in an homogeneous distribution, that is given by $\langle r_i \rangle = (3/4\pi\rho)^{1/3} \Gamma(i+1/3)/(i-1)!$, where $\rho$ is the density of sources. We show in Figure \ref{delt} the amplitude of the dipole and the dispersion obtained in 1000 simulations for two different values of the source density: $\rho = 10^{-5}$ Mpc$^{-3}$, for which the closest source is at a mean distance $\langle r_1 \rangle \simeq 25$ Mpc (solid lines), and  $\rho = 10^{-4}$ Mpc$^{-3}$, for which  $\langle r_1 \rangle \simeq 11$ Mpc (dashed lines). 
For  $\rho = 10^{-5}$ Mpc$^{-3}$ the dipole amplitude rises from about $1\%$ at $E =$ 0.1 EeV to $\sim 2 \%$ at 1 EeV and $\sim 8 \%$ at 10 EeV. 
For  $\rho = 10^{-4}$ Mpc$^{-3}$ an anisotropy smaller by a factor of about 2 results, as many more sources contribute to the total flux  in this  case. 
The results are shown for two values of the amplitude of the turbulent magnetic field, $B= 1$ nG (left panel) and $B= 3$ nG (right panel). A slightly smaller anisotropy amplitude results for the larger turbulent field. At the largest energies, above the GZK cutoff, a steep increase of the anisotropy results as the number of contributing sources decreases. 
Note that  the computations are performed using the stochastic results (Section 4.2) for energies above $E_c$, while below $E_c$ the diffusion solution from Section~3 is used. This explains the small jumps observed  in the plots at $E\simeq E_c$.

One should also keep in mind  that fixing the distances to the sources to the mean expected values and considering equal intrinsic luminosity sources leads to an underestimation of the dispersion in the plots, although the mean values should not be much affected by these approximations.

\begin{figure}[t]
\centerline{\epsfig{width=2.2in,angle=270,file=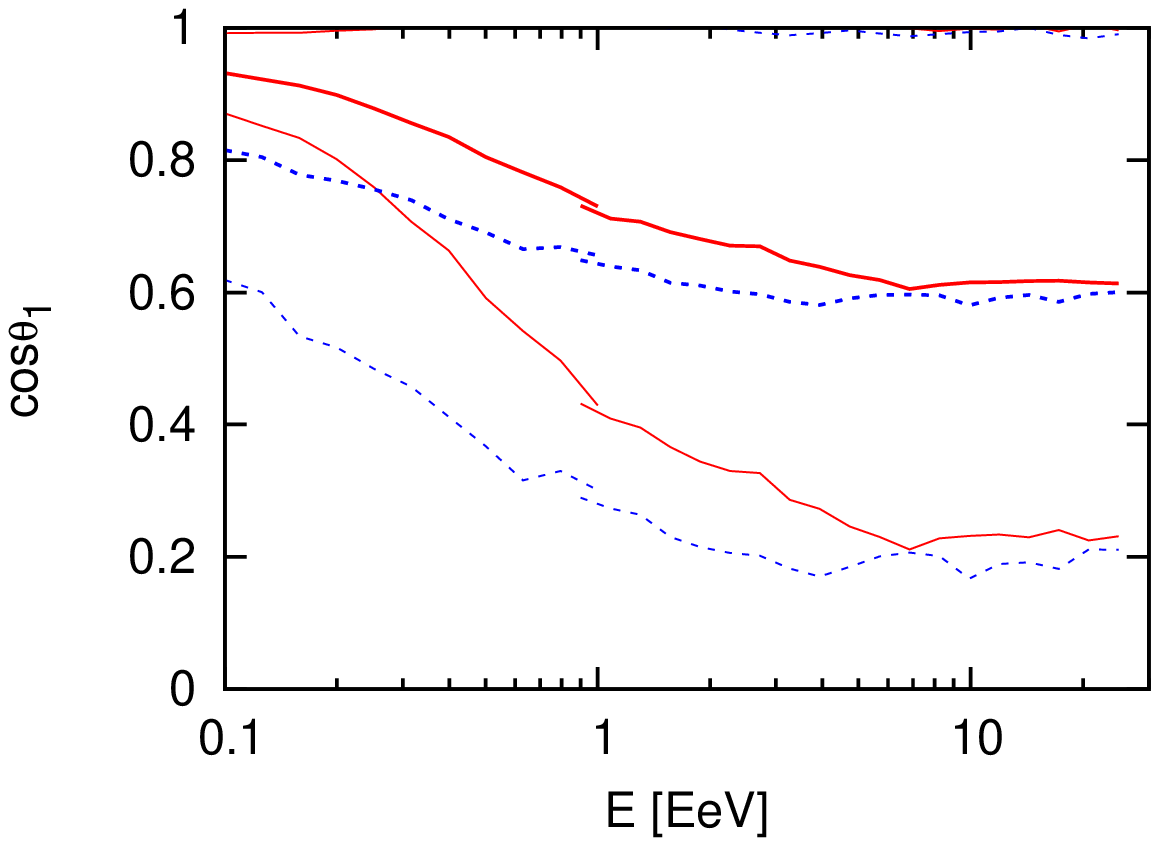},\epsfig{width=2.2in,angle=270,file=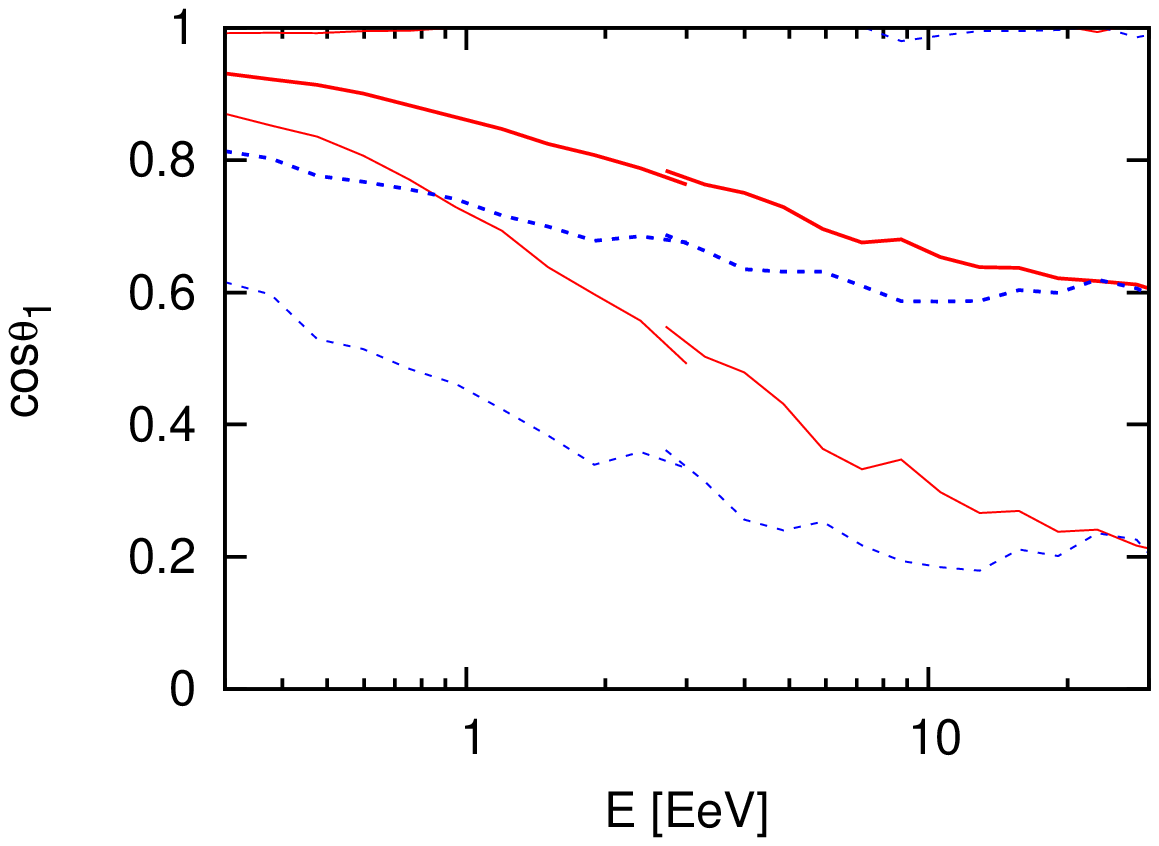}}
\vskip 1.0 truecm
\caption{Mean and dispersion of the cosine of the angle between the directions of the total dipole and that of the closest source as a function of the energy. Left panel: for a turbulent magnetic field of $B= 1$ nG and a density of sources $\rho = 10^{-5}$ Mpc$^{-3}$ (red solid lines) and $\rho = 10^{-4}$ Mpc$^{-3}$ (blue dashed lines). Right panel: same for $B= 3$ nG.}
\label{cosq1}
\end{figure}
We can also wonder how close to the direction towards the closest source is expected to be the direction of the total dipole. The mean and dispersion in 1000 simulations of the cosine of the angle between them is shown in Figure \ref{cosq1}. We see that the mean angle in the case $B= 1$ nG and $\rho = 10^{-5}$ Mpc$^{-3}$ (left panel, solid lines) rises from about $20^{\circ}$ at $E=$ 0.1 EeV to $\sim 45^{\circ}$ for 1 EeV and $55^{\circ}$ for $E > 10$ EeV. These values increase a bit for larger densities and decrease for larger $B$ (right panel).
\begin{figure}[t]
\centerline{\epsfig{width=2.2in,angle=270,file=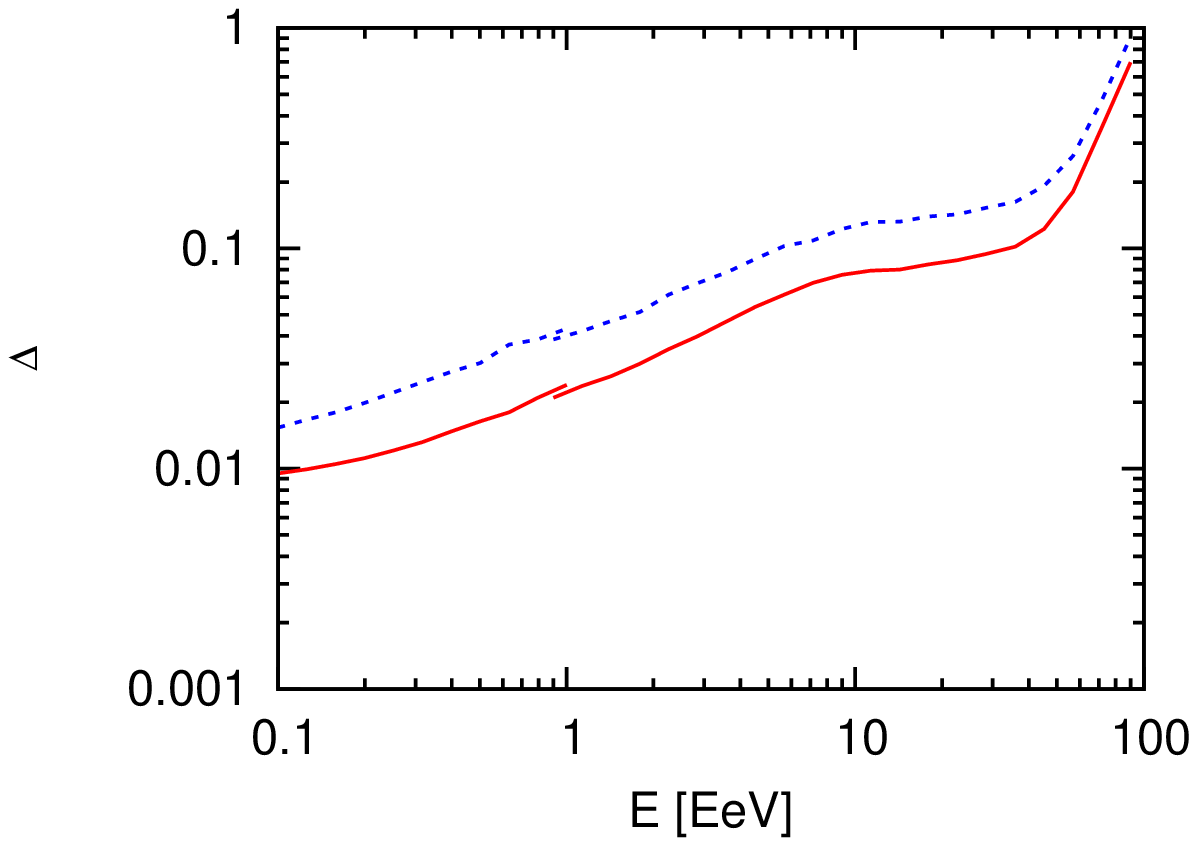}}
\vskip 1.0 truecm
\caption{Mean amplitude of the total expected dipole when local sources within 100 Mpc are distributed like galaxies in the 2MRS catalog (blue dashed lines) considering a density $\rho  = 10^{-5}$ Mpc$^{-3}$ and a turbulent field with $B= 1$ nG. For reference, the red line shows  the expected amplitude for uniformly distributed sources for the same parameters.}
\label{dip2mrs}
\end{figure}

The previous results hold for homogeneously distributed sources. If the sources themselves have instead an inhomogeneous distribution around the observer, in particular with their distribution having  a non-vanishing dipole, a further contribution to the anisotropy is expected. If the local distribution of cosmic ray sources follows the local distribution of matter, actually a non-vanishing dipole is expected. This dipolar component of the matter distribution is indeed known to be responsible for the Local Group peculiar velocity with respect to the rest frame of the CMB, that actually gives rise to the observed CMB dipole. The dipolar component of the mass distribution in our neighborhood has been estimated using different catalogs of galaxies as for example the 2 Micron All-Sky Redshift Survey (2MRS),  showing that the resulting dipole seemingly converges when sources up to a distance $\sim 90$ Mpc are included \cite{2mrs}.

The effect of the local inhomogeneity of the source distribution  in the predicted large scale anisotropies can be included in the simulations by choosing the positions of the sources in our neighborhood from some catalog representing the local distribution of matter.  To describe the local distribution of matter we use a volume limited subsample\footnote{Considering only objects with $d <$ 100 Mpc and absolute magnitude in the K band $M_K < -23.4$} of the 2MRS catalog up to 100~Mpc \cite{hu12}. We have then selected the position of the required number of sources (according to the density considered) from this subsample of  2MRS galaxies. On the other hand, the locations of sources farther away were assumed to be  isotropically distributed. We show in Figure \ref{dip2mrs} the change in the mean dipole amplitude when the inhomogeneous source distribution is considered for a density  $\rho  = 10^{-5}$ Mpc$^{-3}$ and a turbulent field of $B= 1$ nG. An enhancement of the dipole amplitude of about $70\%$ on average is observed with respect to the isotropic case.

\section{Summary and discussion}

We have considered in detail the diffusion of charged particles in turbulent  magnetic fields, obtaining through numerical simulations of the trajectories expressions for the diffusion coefficients. We focused in the computation of the dipolar anisotropies, matching the analytic and numerical results of the 
high energy regime of angular diffusion (quasi rectilinear propagation) with the low energy regime of spatial diffusion.
We illustrated the results for typical  values of $E_c\simeq 1$--3~EeV and  $l_c\simeq 1$~Mpc, showing that the dipole amplitude resulting from  sources with number densities of $10^{-5}$ to $10^{-4}$~Mpc$^{-3}$ are at the level of (0.5--1)\% at 0.1~EeV energies, increasing to (1--2)\% at 1~EeV and up to (3--10)\% at 10~EeV (fig.~4). When the anisotropy in the local (within 100~Mpc) distribution of sources, modeled following the 2MRS galaxy catalog, is taken into account, an increase in the expected dipole amplitude typically by a factor 1.5 to 2 is predicted. In this case this contribution would point in the approximate direction of the motion of the Local Group with respect to the CMB rest frame\footnote{In the rest frame of the CMB the Local Group  moves towards the direction $(\alpha,\delta)=(163^\circ,-27^\circ)$ \cite{2mrs}.}, since it is just the anisotropy in the galaxy distribution that is ultimately responsible for both the proper motion of the Local Group and for the anisotropy in the CR source distribution. These anisotropies are significantly larger than the ones that would result from the Compton-Getting effect \cite{cg} if CRs were isotropic in the rest frame of the CMB, which would be at the level of $\sim 0.6$\% almost independently of the energy.

It should be mentioned that the further deflections of the CRs caused by the galactic magnetic field (mostly by the regular component), not included in this work, would modify the CR dipole amplitude and direction as well as generate higher order multipoles in the arrival directions distribution (see \cite{ha10}). This effect could reduce to some extent  the amplitudes obtained here for energies below few EeV.

Finally, in this work we obtained  the dipolar anisotropies under the assumption of a proton CR composition (which is consistent with the observations at EeV energies and also compatible with HiRes and Telescope Array measurements at higher energies \cite{hiresxmax}). However, if the CR composition were to become heavier above a few EeV, as suggested by the Auger Observatory  measurements \cite{augerxmax}, the anisotropies above the ankle will depend on the details of the actual source composition.

The dipolar amplitudes computed in this work are of interest to interpret  the recent results obtained  by the Auger Observatory hinting at non-vanishing dipolar amplitudes at energies above $\sim 1$~EeV \cite{augerls}, and the possible connection of these results with the non-uniform distribution of nearby sources should be further scrutinized. 

\section*{Appendix I: Turbulent field}

The turbulent magnetic field is modeled as a Gaussian random field with zero mean and root mean square value $B=\sqrt{\langle B^2(x)\rangle}$. This can be described by a superposition of Fourier modes as \cite{gi99,ac99,ha02}

\begin{equation}
B_i({\vec x})=\int \frac{{\rm d}^3k}{(2\pi)^3} B_i({\vec k})
e^{i(\vec{k}\cdot {\vec x}+\phi_i({\vec k}))}\,,
\label{bfourier}
\end{equation}
where the phases $\phi_i({\vec k})$ are random. Under the assumption of isotropic and homogeneous turbulence 
the average of the random Fourier modes verify
\begin{equation}
\langle B_i({\vec k}) B_j ({\vec k'})\rangle = \frac{w(k)}{
k^2}P_{ij}(2\pi)^6 \delta ({\vec k} + {\vec k'})\,.
\label{bibj}
\end{equation}
The projection tensor $P_{ij}=\delta_{ij}-{k_ik_j/k^2}$
guarantees that the field is solenoidal (${\vec \nabla}\cdot {\vec
B}=0$)~. The function $w(k)$ describes the distribution of magnetic energy density on different scales. We consider generic power laws:
\begin{equation}
w(k)=\frac{B^2}{8\pi} k^{-m}\frac{(m-1)(2\pi/L_{max})^{m-1}}{ 1-(L_{min}/ L_{max})^{m-1}}\,,
\label{bk}
\end{equation}
for $2\pi/L_{max}\leq k \leq 2\pi/L_{min}$, and zero
otherwise.   The case of a Kolmogorov spectrum corresponds 
to a spectral index $m=5/3$, and the value $m=3/2$ describes a Kraichnan spectrum. The spectrum is normalized such that
$\langle|{\vec B}({\vec x})|^2\rangle= B^2$.

The correlation length $l_c$ is defined through
\begin{equation}
\int_{-\infty}^\infty {\rm d}l\langle {\vec B}(0) \cdot {\vec B}({\vec
x}(l))\rangle\equiv B^2l_c\,,
\label{lcoh} 
\end{equation}
where the point ${\vec x}(l)$ is displaced with respect to the origin
by a distance $l$ along a fixed direction.  The integral in the lhs of
eq.~(\ref{lcoh}) can be computed using Eqs.~(\ref{bfourier}) and
(\ref{bibj}), and leads to
\begin{equation}
\pi \int_0^\infty \frac{{\rm d} k}{ k}w(k)=\frac{B^2}{8\pi}l_c 
\label{lcoh2}
\end{equation}
which can be used to express $l_c$ in terms of $L_{\rm min}$ and
$L_{\rm max}$ as written in eq.~(\ref{lc}) in the main text. If the spectrum is either  very steep 
$(m\gg 1)$ or  very narrow-band
$(L_{\rm min}\sim L_{\rm max})$, then  $L_c\simeq L_{\rm
max}/2$. If the spectrum is broad-band $(L_{\rm max}\gg L_{\rm min})$ then $l_c\simeq L_{\rm
max}/5$ for a Kolmogorov energy-distribution  while $l_c\simeq L_{\rm
max}/6$ for a Kraichan spectrum.

The numerical realization of the turbulent field used in this work was obtained with the
superposition of $N$ independent modes, with random directions of the
wave-vector ${\vec k}$. The direction of ${\vec B}({\vec k})$ was chosen randomly
in the plane orthogonal to ${\vec k}$, in order to automatically fulfill the condition ${\vec
\nabla}\cdot {\vec B}=0$. The amplitude of
each mode $|{\vec B}({\vec k})|$ was drawn from a Gaussian distribution
with zero mean and root mean square equal to $B$, with $m=5/3$ or $m=3/2$ 
for a Kolmogorov or a Kraichnan spectrum respectively. The wavenumbers of the modes were 
distributed with a constant logarithmic spacing between $k_{\rm min}=2\pi/L_{\rm max}$ and 
$k_{\rm max}= 2\pi/L_{\rm min}$, since more modes are required at larger scales for a better 
approximation to isotropic turbulence. $L_{\rm min}$ was taken as $L_{\rm max}/50$ for high energies, and 
as $r_L/9$ for energies at which this quantity becomes smaller. The number of modes used was larger for larger energies to 
guarantee convergence. The numerical solution of the Lorentz equation in the turbulent field was performed
with a fourth-order Runge-Kutta method.

\section*{Appendix II: Inclusion of energy losses}

The energy losses are described by the coefficients $b=-{\rm d}E/{\rm d}t$, so that the energy loss length can be introduced as
\begin{equation}
\lambda\equiv -c\left(\frac{1}{E}\frac{{\rm d}E}{{\rm d}t}\right)^{-1}=-\frac{cE}{b(E)}.
\end{equation}
In general the losses can be split as $b=b_z+b_{int}$, where $b_z$ results from redshift losses just due to the expansion of the Universe, i.e.
\begin{equation}
b_z=-\frac{E}{1+z}\frac{{\rm d}z}{{\rm d}t}
\end{equation}
with
\begin{equation}
\frac{{\rm d}z}{{\rm d}t}=-H_0(1+z)\sqrt{(1+z)^3\Omega_m+\Omega_\Lambda}.
\end{equation}
The term $b_{int}$ results from the  interactions with the photon backgrounds, mostly the CMB one but in the case of nuclei also the IR and optical/UV ones. 

\begin{figure}[t]
\centerline{\epsfig{width=2.5in,angle=-90,file=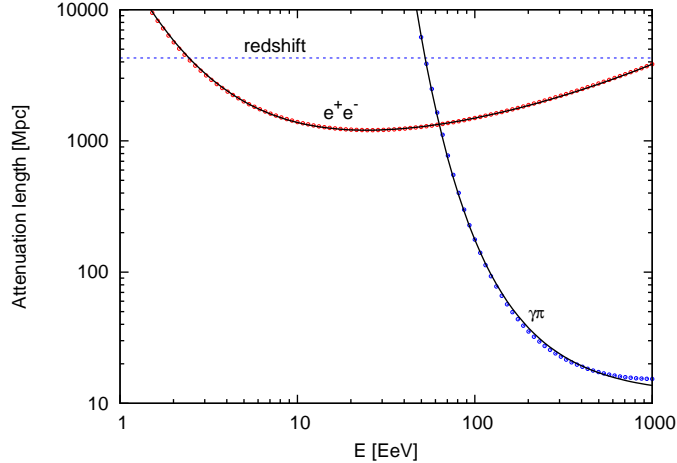}}
\vskip 1.0 truecm
\caption{Proton attenuation length as a function of the energy. Dots are the results obtained in \cite{horizon} while solid lines are the analytic fits introduced in this work. } 
\label{ffits}
\end{figure}

Focusing the attention on the attenuation of the protons, the interactions are just with the CMB and arise from photo-pion processes ($p\gamma\to \pi^{{}+{}}n, {} \pi^0 p$) and from pair production ($p\gamma\to e^{{}+}e^{-}p$), i.e. $b_{int}=b_{\pi N}+b_{e^+e^-p}$. The attenuation lengths can be obtained from the cross sections and inelasticities of the processes, and here we introduce a convenient analytic fit to those results, accurate at the few \% level in the relevant energy ranges ($10^{17}$--$10^{20}$~eV for pair production and $3\times 10^{19}$--$10^{21}$~eV for photo-pion production). 
Introducing the function $F(A,B,C,E)\equiv A\exp(B\,E^C)$, with $E$ in EeV,
we find, fitting the results in \cite{horizon}, that  the present time ($z=0$) proton attenuation length for photo-pion production has the approximate expression
\begin{equation}
\lambda^{z=0}_{\pi N}(E)\simeq F(11.5,686,-1.2,E)\ {\rm Mpc},
\end{equation}
while the proton  attenuation length for  pair production can be expressed as
\begin{equation}
\lambda^{z=0}_{e^+e^-p}\simeq \left[ F(300,4.42,-0.6,E)+F(51,1.61,0.14,E)\right]\ {\rm Mpc}.
\end{equation}
In Figure \ref{ffits} we show the proton attenuation lengths computed in  \cite{horizon} (with dots) and the corresponding analytic fits just introduced (in solid lines).

On the other hand, since the CMB photon density scales as $(1+z)^3$ while its temperature as $T_{CMB}\propto (1+z)$, one can show that for protons
\begin{equation}
b_{int}(E,z)=(1+z)^2b_{int}\left((1+z)E\right).
\end{equation}
The energy $E_g(E,z)$ that a CR had at redshift $z$, given that its present energy at redshift $z=0$ is $E$, is obtained by numerically integrating the equation
\begin{equation}
\frac{{\rm d}E'}{{\rm d}z}=\frac{E'}{1+z}+\frac{(1+z)b_{int}^{z=0}\left((1+z)E'\right)}{H_0\sqrt{(1+z)^3\Omega_m+\Omega_\Lambda}},
\end{equation}
and the change in the energy bin width from the time of production at redshift $z_g$ to the present time is \cite{be06b}
\begin{equation}
\frac{{\rm d}E_g}{{\rm d}E}=(1+z_g)\exp\left\{\frac{1}{H_0}\int_0^{z_g}{\rm d}z\ \frac{(1+z)^2}{\sqrt{(1+z)^3\Omega_m+\Omega_\Lambda}}\ \left.\frac{{\rm d}b_{int}^{z=0}(E')}{{\rm d} E'}\right|_{E'=(1+z_g)E}\right\},
\end{equation}
where the derivative of the energy loss coefficient can be easily obtained from the analytic fits introduced before.

\section*{Acknowledgments}
 Work supported by CONICET and ANPCyT, Argentina.

\end{document}